\documentclass[letterpaper,prl,aps,twocolumn,showpacs,floatfix,superscriptaddress]{revtex4}
\usepackage{epsfig}
\usepackage{bm} 
\usepackage{times}
\usepackage{mathptm}
\usepackage{mathptmx}
\usepackage{sub_JP}

\def\dlap#1{{\setbox0=\hbox{#1}\dimen0=\ht0\divide\dimen0by 1\raise-\dimen0\box0}}
\def\ulap#1{{\setbox0=\hbox{#1}\dimen0
=\ht0\divide\dimen0 
by 1\raise\dimen0\box0}}
\def\nlap#1{{\setbox0=\hbox{#1}\dimen0
=\ht0\divide\dimen0 
by 2\raise-\dimen0\box0}}
\def\figurewithtex#1#2#3#4#5#6#7#8#9{
\def\myunit{1cm}
\setlength{\unitlength}{\myunit}
\def\mywidth{#1}
\def\myheight{#2}
\def\myxorig{7.5} 
\def\myyorig{5}
\def\picoffset{#3}
\def\mytxoff{#4}
\def\mytyoff{#5}

\begin{figure}[ht]\label{fig:#9}
\leavevmode\raise0\unitlength\vbox to\myheight\unitlength{%
\vfill\hbox to\mywidth\unitlength{%
\hfill%
\newdimen\figcenter
\figcenter=\hsize\relax
\divide\figcenter by 2%
\begin{picture}(0,0)(\myxorig,\myyorig)%
\put(0,0){\makebox(21,10)[lb]{\epsfig{figure=#6}}}%
\input{#7}
\end{picture}%
\hfill}%
\vfill}%
\null\vskip\picoffset cm
\caption{#8}
\end{figure}
}%
\renewcommand{\epsilon}{\varepsilon}

\def\HALF{{\textstyle\frac{1}{2}}}

\def\OO{{\cal O}}
\def\real{{\bf R}}
\def\n{{\rm n}}
\def\t{{\rm t}}
\def\d{{\rm d}}
\begin{document}
\title{Temperature Profiles in Hamiltonian Heat Conduction}
\author{Jean-Pierre Eckmann}
\affiliation{D\'epartement de Physique Th\'eorique, Universit\'e de
Gen\`eve}
\affiliation{Section de Math\'ematiques, Universit\'e de Gen\`eve}
\author{Lai-Sang Young}
\affiliation{Courant Institute for Mathematical Sciences, New York University}
\begin{abstract} We study heat transport in the context of 
Hamiltonian and related stochastic models with nearest-neighbor coupling,
and derive a universal law for the temperature profiles of a large
class of such models. This law contains a parameter $\alpha$, and is
linear only when $\alpha=1$. The value of $\alpha$ depends on
energy-exchange mechanisms, including the range of 
motion of tracer particles and their times of flight.
\end{abstract}
\pacs{05.70.Ln, 05.45.-a}
\maketitle
\def\xiL{\xi_{\rm L}}
\def\E{E}
\def\xiR{\xi_{\rm R}}
\def\TL{T_{\rm L}}
\def\TR{T_{\rm R}}
\def\d{{\rm d}}
\def\CA{communicating agent}
\def\ES{energy storing device}
\def\equ#1{(\ref{equ:#1})}
\def\nbf{\bf}
In nonequilibrium physics, the Fourier law is an example of a simple
phenomenological principle whose molecular origin is very hard to
explain. Idealizing homogeneous thin rods and wires with uniform cross-section as 1-D objects, this law says
that heat flux is
proportional to temperature gradient times heat conductivity.
Ever since Fourier's pioneering work \cite{Fourier}, physicists have
tried to derive this law from first principles. 
The current state of the art
is summarized in the excellent reviews
\cite{Lepri2003,Reybellet2000,Casati2003}, which all point to the need
for a deeper theoretical understanding beyond the many existing
models and simulations. In this Letter, we report on new results for
certain types of  Hamiltonian systems and their stochastic realizations.

To study the Fourier law in a Hamiltonian context, the most common 
setting is that of a chain of identical units comprised of
disks, plates, penduli, and the like, 
coupled with short range forces between them.
At its two ends, the chain is coupled to mechanisms 
simulating heat baths maintained at two different temperatures.
Because the components of the chain are identical, one may expect
heat conductivity to be constant along the chain, so that by 
Fourier's law, the temperature profile is linear. 
This seems to be the predominant thinking behind much of the
recent work on Fourier's Law, although nonlinear profiles are known to
occur in other contexts \cite{Hirschfelder}.

In this Letter we point out some simple and
natural mechanisms that lead to various profiles -- both 
linear and nonlinear -- in concrete Hamiltonian models,
and derive a universal law for the profiles of these and
other systems.

{\nbf Summary of results}:

\noindent (A) We consider Hamiltonian models consisting of  
a chain of {\it energy storing devices} (ESD) 
that are fixed in place and coupled to each other.
For such a setup, we show that the temperature profile 
can be linear or nonlinear depending on the nature of the coupling.
More specifically, we assume that
energy exchange in the system is mediated
by {\bf tracers}, which move from ESD to ESD redistributing energy
according to the rules introduced in \cite{Larralde2002,Larralde2003}.
We find that {\em the profile is linear if energy transfer is carried
out by a single tracer that moves freely along the chain, 
whereas heat conductivity is temperature-dependent (and hence
the profile is nonlinear) if the tracers are confined to specific 
regions.}

\noindent (B) Our second result is
a {\em universal law} that holds for very general 
coupled chains of Hamiltonian or stochastic systems with {\it nearest-neighbor} 
interaction (including those considered in (A)). We show that 
as the number of constituent cells 
goes to infinity, the stationary temperature profile 
is given by
\begin{equation}\label{equ:1}
  T(x)= \bigl(\TL^\alpha  +(\TR^\alpha -\TL^\alpha )x\bigr)^{1/\alpha }~
\end{equation}
where $\TL$ and $\TR$ are the temperatures imposed 
at the left and right ends,
$x$ is the coordinate along the system (normalized to
$x\in[0,1]$), and $\alpha$ is a constant. The form of this
 law does not depend on details of the system
(precise conditions are given later). The value of $\alpha$
in (1), however, depends on the nature of the coupling.

In the case of locked-in tracers, {\em i.e.}, tracers confined to 
the regions between neighboring ESD, the value of $\alpha$
depends on their {\it time of flight}. In particular, $\alpha = 
\frac{3}{2}$ if the energy is purely kinetic.
In the case of a single tracer moving freely along the chain, $\alpha =1$
independent of its time of flight.
We show the results above for 
stochastic realizations of the models in (A), and explain
why one should expect these results to carry over to their
Hamiltonian counterparts.

Our primary concern in this Letter is the
temperature profile  (TP). The existence of 
local thermal equilibrium (LTE), which is
important for the definition of temperature, will be proved 
in \cite{jpls} in the infinite volume limit for one of 
the models treated here. 
(For results on LTE for other models, see \cite{Dhar1999,
Galves1981,Kipnis1982}.)

Before embarking on specifics, we note again that in our models,
we distinguish between {\em communicating agents}
(CA), which in our case are tracer particles, and ESD, which
 in our case are turning disks with fixed centers (we emphasize that the
ESD are {\em   not} infinite reservoirs).
In a real-world conductor, the difference between CA and ESD
is often blurred, but for a good theoretical understanding 
it is useful to keep them separate.
These concepts are distilled from the following beautiful model:

\begin{figure}
\leavevmode\input fig1.inp
\includegraphics[width=.4\textwidth]{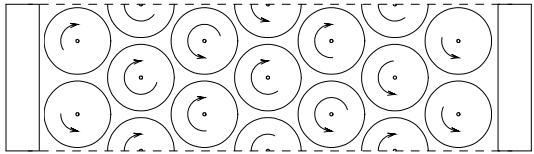}
\caption{\label{fig:larralde}A typical arrangement of disks in the
MLL  Model. The simulations in 
\cite{Larralde2003} were done with 2 rows and periodic boundary conditions in the
vertical direction, and zig-zag reflecting walls of temperature
$\TL$, resp.~$\TR$, at the two ends. The tracer particle is not shown.}
\end{figure}


{\nbf The MLL Model} \cite{Larralde2002,Larralde2003}: 
We describe this model in some detail, as it contains the basic ingredients 
of the models in (A).
The MLL Model is purely Hamiltonian,
  and very careful simulations show that the Fourier law  holds.
  The system consists of an
  arrangement of $N$ disks of radius 1 placed as in 
Fig.\ref{fig:larralde}, 
  and a little point particle of mass 1 (the tracer) which
  wanders around the playground $\Omega$ (the physical space
occupied by the system minus the disks), bouncing off the disks
  \footnote{More than one tracer is sometimes used 
  in \cite{Larralde2003}, but we will limit ourselves
to the single tracer case.}.
While Fig.\ref{fig:larralde}
suggests a Lorentz gas \cite{Lorentz1905}, there is a
  crucial difference here: Each disk is ``nailed down'' in its center,
  around which it turns freely. The state of the system is described by
$x= (\omega _1,\dots,\omega _N,q,v)$ 
where $\omega _{i}$ is the
angular velocity of disk ${i}$, $q\in\Omega $ is the position of the
tracer, and $v$ is its velocity. When the
tracer collides with a disk, the rule of interaction is that of
``sticky reflection": Suppose the angular velocity of the disk being hit
is $\omega $, and $v_\n$ and  $v_\t$ are the
normal, resp.~tangential, component of $v$ relative to the impact point.
Then the values of $v$ and $\omega $ after the collision
are given by the energy and angular momentum conserving law
\begin{eqnarray}\label{equ:exchange}
v'_\n\,=\,-v_\n~,\quad
v'_\t\,=\,v_\t-\frac{2\epsilon}{1+\epsilon }(v_\t-\omega )~,\nonumber\\
\omega' \,=\,\omega +\frac{2}{1+\epsilon }(v_\t-\omega )~.
\end{eqnarray}
Here, $\epsilon \in\real^+$ is proportional to the moment of inertia
of the disk
divided by the mass of the tracer and the square of the radius of the disk.
\cite{Larralde2003} treats mostly the case $\epsilon=1$, where
$v'_\t=\omega$ and $\omega'=v_\t$, {\it i.e.}, the two quantities are
simply exchanged. 
Of particular interest to us is the case $\epsilon\ll 1$, which
from the tracer's point of view resembles the classical 
Lorentz gas.

{\em Remark.} The MLL Model, as well as the Hamiltonian models
 we will describe later, have the 
following important property: Since there is only a hard-core
potential, the time evolution of the system is {\em rescaled} by
$\sqrt{\lambda }$
when the energy of the particle and the disks are rescaled by 
$\lambda
$. In this respect, the model in \cite{Larralde2003} 
is very different from
models such as the ding-a-ling and ding-dong models
\cite{Casati1984,Prosen1992,Garrido2001}. 
Most importantly, the energies of the tracer and the disks
alone determine the time-of-flight of the tracer: It does not depend on
the history (as it would in many models considered so far
\cite{Casati1984,Prosen1992,Garrido2001}). 

We still need to say what happens when the tracer hits one of
the ends. In \cite{Larralde2003}, many variants are considered, 
but for our purpose, the following process is assumed:
When the tracer hits one of the ends, it exits the system, 
and a new tracer is injected into $\Omega$ to take its place.
The new tracer enters $\Omega$ at the point of exit of the old one.
Its direction is arbitrary, and its speed is given by 
the Maxwell distribution for the temperature of the end in question.


{\nbf Introducing the models studied in this Letter}: We now introduce
two classes of models that have the same basic setup of tracers 
and turning disks 
(and the same rules of interaction) as in the MLL Model. 
However, the  configurations of disks and tracers 
in these two models are chosen to give rise  to 
two conceptually very different modes of transport.

{\nbf Model I (Wandering tracer)}: 
A single tracer wanders along a chain of
boxes separated by walls with a tiny hole that allows the tracer 
to pass 
between adjacent boxes. Deep inside each box is a turning disk 
surrounded by
many fixed disks. The turning disk serves as ESD, while the fixed disks 
are {\em bona fide} Lorentz scatterers, which serve to randomize
 the angles of incidence in collisions between the
tracer and the turning disk, leading to the exchange of
a random portion ({\em i.e.}, the tangential component) of the energy 
of the tracer. The smallness of the holes in the separating walls 
keeps the tracer in each box for a long period. This together with the
chaotic action of the Lorentz scatterers ensures that the tracer  
is equally likely to exit the box from either side.

{\em Stochastic realization}: We model this system on a 1-D lattice
with  $N$ sites. There is a random variable $\xi_{i}$ representing
 the energy at site $i$, ${i}=1,\dots,N$, with values in
$[0,\infty )$. The heat baths at the ends are modeled by stochastic
variables $\xiL$ and $\xiR$ which take values in $[0,\infty )$ with a
distribution $\TL\exp(-\xiL/\TL)$, resp.~$\TR\exp(-\xiR/\TR)$ (the
Boltzmann constant $k_{\rm B}$ being set to 1). 
We identify $\xiL$ with the variable $\xi_0$, and $\xiR$ with
$\xi_{N+1}$.  There are two
more random variables, $\eta$, to be
  thought of as the energy of the tracer, and ${i}$, which gives
 the location of the tracer at any given time. 
We assume that when
the tracer is at site ${i}$, it interacts with $\xi_{i}$.
At a site ${i}$, ${i}\notin \{0,N+1\}$, the action is as
follows:
There is a clock which rings with rate $f$, for example
$f=\eta^{-1/2}$ or 
$(\eta+\xi_{i})^{-1/2}$,  representing the time it takes for the 
tracer to make its way around the ${i}$th box.
When the clock rings, the following mixing of energies 
takes place \footnote{Similar stochastic rules for mixing energies
are used in  \cite{Galves1981,Kipnis1982}.}:
Choose a random variable $p$ with uniform distribution in
$[0,1]$. 
Then, 
\begin{equation}\label{equ:etaxi}
\eta'=p(\xi_{i}+\eta),\quad
\xi'_{{i}}= (1-p)(\xi_{i}+\eta)~.
\end{equation}
If ${i}=0$, then $\eta$ is replaced by a value chosen from the
exponential distribution for the temperature $\TL$. The
rule at the right end (${i}=N+1$) is similar.
After these operations, the tracer jumps with probability 
$\HALF$ to ${i}-1$ or
${i}+1$ -- except when it is at the ends, in which case it stands still
with probability $\HALF$ or moves into position 1 (resp.~$N$) from
the boundary.
{\em We prove in \cite{jpls} that for all such models, the TP
is linear and the distributions of the $\xi_i$ satisfy LTE.} 

\begin{figure}
\leavevmode\input fig2.inp
\includegraphics[width=.4\textwidth]{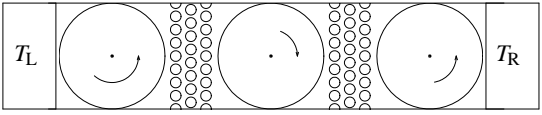}
\caption{\label{fig:fig2} A sketch of Model II, when 
it is made more chaotic. Between the
  rotating disks, there are disks serving as Lorentz scatterers. The
  tracers are not shown. The horizontal walls are reflecting.}
\end{figure}

{\nbf Model II (Locked-in tracers)}: In this model there is
 a channel one-disk wide, with reflecting boundaries, 
and with turning disks located
at fixed distances apart. These disks turn freely, but they block 
the channel
completely, separating it into individual cells. Inside each cell 
is a single
tracer, which moves back and forth, transferring energy between the 
turning disks that border the cell \footnote{A version of this model was
suggested to us by D. Ruelle.}. Here one can  assume the tracer hits
the two turning disks alternately, or, to further randomize the 
situation, one can add a number of 
fixed disks in each cell as illustrated in Fig.~\ref{fig:fig2}.
After hitting one turning disk, the tracer then ``gets lost" 
in this array of
Lorentz scatterers, to emerge at some random moment to hit 
the turning disk
at either side with equal probability. In both cases, the time-of-flight
of the tracer between hitting turning disks depends only on the speed 
of the tracer (and not on the state of the disks). 

\begin{figure}
\leavevmode\input fig3.inp
\includegraphics[width=.4\textwidth]{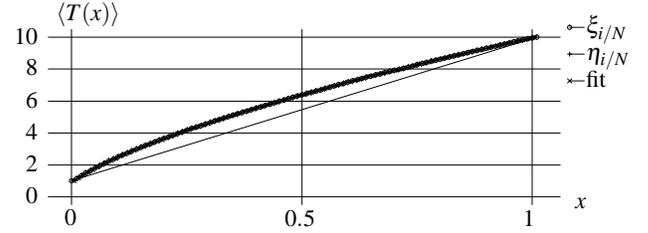}
\caption{\label{fig:fig3}The
mean values of $\xi$ and $\eta$, as a function of $x=i/N$,
with $\TL=1$, $\TR=10$,
$N=100$,
averaged over $2\cdot
10^9$ exchanges of energy according to Model II. Superposed is the
theoretical curve of Eq.\equ{1}. Note that the temperature
profile is {\em not} linear and that its curvature is more pronounced
at the cold end. Here, $\alpha =\frac{3}{2}$.
We have obtained the same profile for many variants of Model II.}
\end{figure}

{\em Stochastic realization}: We have $N$ sites on a 1-D 
lattice as in the previous stochastic model, with $\xi_i$ 
representing the energy of the turning disk at site $i$. 
 In this model, however,
there  is one independent variable $\eta_{i}$ for each
pair $({i},{i}+1)$,
  ${i}=0,\dots,N$, representing the energies of the tracers.
Each site is equipped with an (independent) clock 
which rings at an exponential rate proportional to $\eta_{i}^{-1/2}$. 
When this clock rings, an exchange of energy 
involving $\eta_{i}$ takes place. As in the Hamiltonian model,
one may assume $\eta_{i}$ exchanges energy 
alternately with $\xi_{i}$ and $\xi_{{i}+1}$, or, 
$\eta_{i}$ chooses with probability $\HALF$ its left or right
partner ({\it i.e.}, $\xi_{i}$ or $\xi_{{i}+1}$), and performs the usual
mixing: For example, if $\xi_{i}$ has been chosen,
then
\begin{equation}\label{equ:mixall}
\eta_{i}'= p(\xi_{i}+\eta_{i})~, \quad
\xi'_{{i}}= (1-p)(\xi_{i}+\eta_{i})~.\\
\end{equation} 
When the clock at site $0$ rings, $\eta_0$ is replaced 
by a value chosen from the
exponential distribution of temperature $\TL$ as before. The
rule at the right end (${i}=N+1$) is similar.
{\it Numerical simulations show clearly profiles deviating
strongly from linearity. They are in perfect agreement with the 
value of $\alpha = \frac{3}{2}$ predicted by theory}
(see Fig.\ref{fig:fig3} and the sketch of argument below).

The qualitative shape of the nonlinear profile can be 
understood easily by considering 3 successive sites, 
say $\xi_{i-1}$, $\xi_i$, and $\xi_{i+1}$. Since $\xi_{i-1}<
\xi_{i+1}$, $\eta_{i+1}$ rattles faster than $\eta_{i}$ 
(the rate being given by $\eta^{-1/2}$). Thus
$\xi_{i}$ equilibrates more often, and hence better, with $\xi_{i+1}$ than
with $\xi_{i-1}$, which explains the concavity of the TP. 

We next present some details of the theoretical arguments for the
results above \cite{jpls}:

{\nbf Reduction from Hamiltonian
to stochastic models}: As dynamical systems, the Hamiltonian
models above are very chaotic. With $\varepsilon$ in the MLL Model taken
to be $\ll 1$, the dynamics are close to those in 
billiards of Sinai type and Lorentz gases.
This chaotic behavior is used to induce a strong 
``memory loss'' for the tracer particles. Fast correlation decay
\cite{young1998}
ensures that the tracer ``forgets'' from which side it enters
a site. This justifies our assumption in
the stochastic model that the fraction of energy exchanged 
is independent from one step to the next.

{\nbf Range of validity of (1).} We consider a coupled chain of length
$N$ in a steady state, and let $E_i$ be the mean temperature at
site $i$. We assume (\i) translation invariance of the model; 
(\i\i) $E_i$ is determined by $E_{i+1}$ and $E_{i-1}$; (\i\i\i) 
energy-scale invariance, meaning if $E_{i \pm1}$ are multiplied 
by $\lambda$, then $E_i$ is also multiplied by $\lambda$; 
(\i v) as $N \to \infty$, the TP tends to a smooth function. 
It is not hard to show that (\i)--(\i v)
imply that the limit stationary density satisfies a second
order differential equation  (namely \ref{equ:22}), the
solution of which with boundary conditions 
$T_L$ and $T_R$ is (1). 

We demonstrate how to derive this law -- and compute the constant
$\alpha$ at the same time --  for Model II:

{\nbf Computing the temperature profile in Model II.}
Consider 3 
successive sites. For illustration we assume (i) each tracer 
visits alternately the left and right disks and (ii)
the mixing of energies at each collision is
exactly half-and-half. The stationarity condition means that the speed
of the tracer is equilibrated as well. Let the mean energy of the left
tracer  be $\eta_{-,\to}$ as it heads toward site ${i}$ and 
$\eta_{-,\leftarrow}$ as it goes away from it. 
By the rule of mixing, we have
$\eta_{-,\to}=(\eta_{-,\leftarrow}+E_{{i}-1} )$,
$\eta_{-,\leftarrow}=(\eta_{-,\to}+E_{{i}} )$, leading to
$\eta_{-,\to}\,=\,(2E_{{i}-1}+E_{{i}})/3$,
$\eta_{-,\leftarrow}\,=\,(E_{{i}-1}+2E_{{i}})/3$.

We assume the speed of the tracer is $E^\gamma$ when $E$ is its
energy.
The value of $\gamma $ for Model II as described above (and 
many other models
without potential) is $\gamma=\HALF$ 
(since the energy is purely kinetic) while for potential
interactions,
the time is given by an integral of the form 
$\int\d q\,(E-V(q))^{-1/2}$,
which for large $E$ and $V(q)\sim|q|^m $ behaves like
$\OO(E^{-1/2+1/m })$, so that $\gamma =\HALF-\frac{1}{m} $.
Fixing $\gamma $, the average time for a round-trip of the tracer
between sites ${i}-1$ and ${i}$ is
$
\tau _-\,=\,\eta_{-,\to}^{-\gamma} +\eta_{-,\leftarrow}^{-\gamma}~,
$
and the rate at which site ${i}$ gets information from the left
is the inverse of this quantity. An entirely analogous reasoning
applies on the ``$+$'' side.
{}From the stationarity condition, we get
\begin{equation}\label{equ:condition}
E_{i}\,=\,\frac{\tau_- ^{-1} (E_{i}+\eta_{-,\to})/2 +
  \tau _+^{-1}(E_{i}+\eta_{+,\leftarrow} )/2}{
\tau _-^{-1}+\tau _+^{-1}}~.
\end{equation}
Perform a perturbative analysis at the point $x={i}/N$ where $N$ is 
very large. Then, to second order in $\epsilon =1/N$,
$
E_{{i}\pm1}\,=\,T(x)\pm\epsilon T'(x)+\HALF \epsilon ^2 T''(x)~,
$
and \equ{condition} leads to
$$
t=T(x) +\epsilon ^2 \frac{T''(x) T(x) +  \gamma \bigl(
  T'(x)\bigr)^2}{4T(x)}+\OO(\epsilon ^4)~.
$$
Since $E_{i}$ must be equal to $T(x)$, we find
\begin{equation}\label{equ:22}
T''(x)T(x)\,=\,-\gamma \bigl(  T'(x)\bigr)^2~,
\end{equation}
the solution of which with boundary conditions $T(0)=\TL $ and
$T(1)=\TR$  is Eq.\equ{1} with $\alpha =1+\gamma $. Thus for
Model II, $\gamma=\HALF$ and $\alpha
=\frac{3}{2}$. One also checks that the energy flux is given by $T'(x)
\sqrt{T}(x)$ (which is constant along the profile, but {\em not}
proportional to the temperature difference).

{\em Generalization.} Note that when $\gamma =0$, {\it i.e.}, when 
the rate at which information is exchanged is independent of energy,
then $\alpha =1$, which indeed gives a linear TP. Note also that
our derivation is quite general: if 
$E^{\gamma}$ is replaced by $1/F(E)$, the profile is given by
$T''(x)F(T(x))\,=\,\bigl(  T'(x)\bigr)^2 F'(T(x))$.

{\em Remark.} Many authors have done careful simulations of
  models that are close to Model II, and
  have observed linear TPs. It should be noted that the 
profiles predicted 
  by \equ{1} are very close to linear when $\TR/\TL$ is not far from $1$.
  Our theory predicts that deviations from
  linearity become more prominent  with the increase of $\TR/\TL$.


{\nbf Linear profile and LTE for Model I.} The reasoning  above
is not valid for Model I, for here it is a single tracer that 
is responsible for all transmission of information to all sites.

In the discussion below, we  refer to the
stochastic realization of Model I as a
{\it variable-time model}, in the sense that the tracer lingers
in each site for variable time periods depending on the
function $f$. 
We introduce an associated {\it fixed-time model}, obtained
by neglecting the waiting time above, {\it i.e.}, all rules are 
as in the variable-time model except that
the tracer jumps at fixed time intervals. 

{\em Linear profile.} Consider first the fixed-time model. 
Here the tracer performs a
standard random walk on $\{1,2, \dots, N\}$ (except at the ends).
It is easy to see that it spends an equal amount of time at each
of the $N$ sites. Indeed for a fixed site $i$, the tracer comes
from the left and the right {\it exactly} the same number of
times over any period, for if the tracer heads left from site $i$, 
then it can only return from the left; and similarly for the right.
Reasoning as above, we see that in
the stationary measure,  $E_i$  is
affected equally by $E_{i-1}$ and $E_{i+1}$, hence the TP is
linear, {\it i.e.}, $\alpha=1$.
Similar reasoning applies to the variable-time model.
Note that the mixing rule, 
and hence the outcome of the mixing process,
does not depend on the waiting time.

{\em LTE}. Methods similar to those in \cite{Galves1981,Kipnis1982} 
with a different
dual XXX give LTE for the fixed-time model. 
While there is a correspondence between the sample paths for the 
variable-time model and its fixed-time counterpart, the varying 
durations the tracer spends at each site affects nontrivially 
the stationary measure. Proving canonical distributions for the
model of interest, {\it i.e.}, the variable-time model, 
requires further control of correlations \cite{jpls}.
Our proof applies to any function $f$ 
that is integrable and {\it local}, meaning it depends only on the 
variables $\xi_i$ and $\eta$ when the tracer is at site $i$
(which is the case here).

{\it Remark}: Even though this LTE result is for the
stochastic realization of Model I, we have incorporated into our setting 
the notions of {\em trajectories} and {\em velocities} of
particles, two of the key ingredients in Hamiltonian dynamics.


{\nbf Conclusion.} (1) We have pinpointed some simple and very natural
mechanisms responsible for both linear and nonlinear profiles in
homogeneous conductors modelled by Hamiltonian systems. 
Our results show in particular that when their ranges of motion are 
restricted, interacting particles (or springs) acquire 
energy-dependent speeds.
(2) We have derived an exact formula -- a universal law -- for the 
energy profiles of very general Hamiltonian and stochastic chains with 
nearest-neighbor interactions. When nonlinear, this law predicts how
deviation from linearity increases with the quotient $\TR/\TL$.
(3) Finally, the underlying causes for nonlinearity that we have
identified clearly go beyond the models studied here. They suggest 
that the presence of some (weak) nonlinear effect may be 
a more common phenomenon than recognized when very disparate 
temperatures are imposed at the two ends of a $1$-D system.

{\nbf Acknowledgments.} We have profited from illuminating discussions
with C. Mej\'\i{a}-Monasterio, D. Ruelle, L. Rey-Bellet, O. Lanford,
Y. Avron and many others.
JPE wishes to thank the Courant
Institute and IHES for their kind hospitality.
This research was partially supported by the Fonds National Suisse and
NSF Grant \#0100538.

\bibliography{refs}

\end{document}